\documentclass[epj]{svjour}

\usepackage{graphics}
\usepackage{epsfig}

\begin{document}

\title{Variational charge renormalization in charged systems}

\author{Roland R. Netz$^+$ and Henri Orland$^*$}

\institute{
$^+$Sektion Physik, Ludwig-Maximilians-Universit\"at, 
             Theresienstr. 37, 80333 M\"unchen, Germany\\
$^*$Service de Physique Th\'eorique, 
CEA-Saclay, 91191 Gif sur Yvette, France}

\date{Received: date / Revised version: date}

\abstract{We apply general variational techniques to the problem 
of the counterion distribution around highly charged objects
where strong condensation of counterions takes place.
Within a field-theoretic formulation using a fluctuating
electrostatic potential, the concept of surface-charge
renormalization is recovered within a simple one-parameter
variational procedure. As a test, we reproduce the 
Poisson-Boltzmann surface potential for a single charge planar
surface both in the weak-charge and strong-charge regime.
We then apply our techniques to non-planar geometries
where closed-form solutions of the non-linear Poisson-Boltzmann
equation are not available. In the cylindrical case, the Manning
charge renormalization result is obtained in the limit
of vanishing salt concentration. However, for intermediate
salt concentrations a slow crossover to the
non-charge-renormalized regime (at high salt) is
found with a quasi-power-law behavior which helps to
understand conflicting experimental and theoretical results
for the electrostatic persistence length of polyelectrolytes.
In the spherical geometry charge renormalization is only found
at intermediate salt concentrations.
}

\maketitle

\section{Introduction}

The behavior of charged systems has attracted renewed interest in the last
few years\cite{Holm}. One of the topics of interest is centered around
the effects of multi-valent counterions, in which case correlations
between ions become important and mean-field type theories break down. 
In such situations it is known experimentally and theoretically that 
equally charged bodies can attract each other. But even for monovalent ions
and moderately to highly charged surfaces many open questions remain:
In such situations the coupling parameter (which measures to which extent
corrections to the mean-field or saddle-point solution are important)
is not very high, but the surface potential can exceed thermal energy 
by far such that non-linear effects become important. The complication
in this case is that the mean-field or Poisson-Boltzmann equation is 
a non-linear differential equation which in the presence of salt ions
can be solved in closed form only
in the planar geometry\cite{gouy}. In the cylindrical and 
spherical geometry a number of different approximations have been proposed,
which all more or less agree on the fact that counterions condense in the vicinity
of the charged surface such as to reduce its effective surface 
charge\cite{Manning2,Pincus}. 
The original Manning condensation argument states that the
effective charge of a charged cylinder is maintained at a constant level
of one charge per Bjerrum length\cite{Manning}. 
The prescriptions for calculating the 
effective charge are numerous and so are the predictions for the precise
value of the effective charge as a function of external parameters such as
salt, temperature, pH and so on\cite{Belloni,Levin,Deserno}. 

In this paper we show how to approach the problem using a field-theoretic
formulation of a variational theory, within which the effective surface
charge is treated as a variational parameter. There are in principle two
different approaches, one is based on the standard density functional
theory, within which the counterion distribution is variationally 
determined. This approach is clumsy for more complicated geometries and
in principle is difficult to generalize such as to include strong-coupling
(i.e. deviations from mean-field) effects. We therefore developed an alternative
approach based on an exact field-theoretic formulation using the fluctuating
imaginary electric potential.
The surface charge is used as a variational parameter and the variational
field-theoretic action is Gaussian.
This is probably the most unambiguous definition of the renormalized
or effective surface charge, since no additional assumptions
(other than that the variational Hamiltonian is Gaussian) is used.
As a trivial byproduct, the effective interaction between charges
is at large distance of the Debye-H\"uckel type with renormalized 
effective charges.
Our scheme in principle allows to go beyond  mean-field,
but we do not pursue such studies in this paper.
As we show in the appendix, the standard Gibbs variational principle can be
reformulated as a perturbative variational scheme which can be systematically
improved by a perturbation series (which we also leave as a problem for the future).
We test our approach for the planar charged wall, for which we essentially exactly
reproduce the Poisson-Boltzmann result.
We then tackle the charged cylinder. The effective cylinder charge interpolates
continuously between the unrenormalized limit (at high salt concentrations)
and the Manning charge-renormalized limit (at low salt concentrations), in agreement
with previous results\cite{Levin}.
However, this crossover occurs for fully charged polyelectrolytes over 
4 orders of magnitude of the salt concentration, and therefore it is important
to take this slow crossover into account. For the controversial debate
on the electrostatic persistence length it means that the standard Odijk-Skolnick-Fixman
result\cite{Odijk,Skolnick,Joanny}, 
obtained on the linear (Debye-H\"uckel) level, is modified 
by the salt dependent effective polymer charge, which changes the 
salt dependency from $\ell_{\rm OSF} \sim \kappa^{-2}$ to 
$\ell_{\rm OSF} \sim \kappa^{-1.4}$, closer to experimental values\cite{Forster}.
For the charged sphere the resultant behavior is different: Here charge
renormalization takes place only at intermediate salt concentrations 
and both for high and low salt concentrations the standard Debye-H\"uckel potential
with unrenormalized charges is valid\cite{Belloni}.
Our method can be easily generalized to more complicated geometries and situations.

\section{The model}

The partition function of a system consisting of 
$N_+$ positively charged and $N_-$ negatively charged
ions of the same valency $q$, interacting with an arbitrary 
charge distribution $\sigma({\bf r})$, can be written as
\begin{eqnarray} \label{part1}
&& {\cal Z  }= 
\frac{1}{N_+!} \prod_{j=1}^{N_+} \left[ 
\int \frac{{\rm d} {\bf r}^+_j}{\lambda_t^3} \Omega( {\bf r}^+_j) \right]
\frac{1}{N_-!} \prod_{j=1}^{N_-} \left[ 
\int \frac{{\rm d} {\bf r}^-_j}{\lambda_t^3} \Omega( {\bf r}^-_j) \right]
\nonumber \\ &&  \exp \left\{
-\frac{\ell_B}{2} \int {\rm d}{\bf r} {\rm d}{\bf r}'
[q \hat{\rho}_+({\bf r})- q \hat{\rho}_-({\bf r}) + \sigma({\bf r})] 
v({\bf r}-{\bf r}')  \right. \nonumber \\ && \left.
[q \hat{\rho}_+({\bf r}')- q \hat{\rho}_-({\bf r}') +\sigma({\bf r}')] 
 +\frac{q^2 \ell_B (N_++N_-) }{2} v(0)
  \right\}.
\end{eqnarray}
The pair potential $v({\bf r})$ denotes the Coulomb interaction,
$v({\bf r})=1/r$, but the first part of our discussion is rather general
and independent of the specific form of $v({\bf r})$.
The self-energy term in Eq.(\ref{part1}),  proportional to
$q^2 \ell_B v(0)/2$, simply subtracts the diagonal term in the
double integral. It will be later shown to renormalize the 
chemical potential (or fugacity) in a rather trivial way.
The Bjerrum length $\ell_B= e^2/4\pi \varepsilon k_BT$ 
is the distance at which 
two unit charges  interact with thermal energy $k_BT$.
The length $\lambda_t$ is the thermal wavelength, 
which results from integrating out kinetic degrees of freedom,
the precise value of which is unimportant.
The function $\Omega({\bf r})$ takes into account the presence
of hard walls and restricts the positions of ions to an appropriate
region in space, as will become more clear when we consider explicit
examples in the following sections.
The counterion density operators are
\[ \hat{\rho}_+({\bf r}) = \sum_{j=1}^{N_+} \delta({\bf r}-{\bf r}_j^+)\]
for positively charged ions
with a similar expression for the negative ions.
At this point, there is not much one can do with the 
partition function, 
mainly because the density operators $\hat{\rho}_+$ and
$\hat{\rho}_-$ enter in a quadratic fashion. 
The main step in deriving a field theory consists in introducing a
unit operator which couples to the density operator $\hat{\rho}_+$,
\begin{eqnarray} \label{delta}
1 &=& \int {\cal D} \rho_+ \; \delta(\rho_+ - \hat{\rho}_+ ) 
\nonumber \\ &=&
\int {\cal D} \rho_+ {\cal D} \psi_+ \exp \left\{
\imath \int {\rm d} {\bf r} \psi_+({\bf r})
[\rho_+ ({\bf r}) - \hat{\rho}_+ ({\bf r})] \right\},
\end{eqnarray}
(and a similar expression for the density of negative ions),
where we used the integral representation of the delta function.
This unit operator can be introduced into the partition function. 
Once introduced, it allows us to replace
density operators $\hat{\rho}_+ ({\bf r})$ and $\hat{\rho}_- ({\bf r})$ 
by the corresponding fluctuating density fields $\rho_+ ({\bf r}) $ 
and $\rho_- ({\bf r}) $ and the configurational integrals over the 
ion positions can be performed.
As a result, the partition function now  reads
\begin{eqnarray}
\label{part3}
&& Z  =
\int {\cal D} \rho_+  {\cal D} \rho_- {\cal D} \psi_+ {\cal D} \psi_-
\nonumber \\ && \exp \left\{ -
\frac{\ell_B }{2} \int {\rm d} {\bf r}
 {\rm d} {\bf r}' [q\rho_+({\bf r})-q\rho_-({\bf r})+\sigma({\bf r}) ]
v({\bf r}-{\bf r}')  \right. \nonumber \\ && \left.
[q\rho_+({\bf r}')-q\rho_-({\bf r}') + \sigma({\bf r}')] 
 \right. \nonumber \\ && \left.
+\imath \int {\rm d} {\bf r} \left[ \psi_+({\bf r}) \rho_+ ({\bf r})
+ \psi_-({\bf r}) \rho_- ({\bf r}) \right]
\right\} \nonumber  \\ &&
\frac{1}{N_+ !} \left[ {\rm e}^{q^2 \ell_B v(0)/2}
\int \frac{{\rm d} {\bf r}}{\lambda_t^3} \Omega({\bf r})
{\rm e}^{- \imath \psi_+({\bf r}) } \right]^{N_+} \nonumber  \\ &&
\frac{1}{N_- !} \left[ {\rm e}^{q^2 \ell_B v(0)/2}
\int \frac{{\rm d} {\bf r}}{\lambda_t^3} \Omega({\bf r})
{\rm e}^{- \imath \psi_-({\bf r}) } \right]^{N_-}.
\end{eqnarray}
The partition function becomes simpler upon 
transformation to the grand-canonical ensemble 
according to 
\begin{equation}
\label{part4}
Z_\lambda = \sum_{N_+,N_-=0}^{\infty} \lambda_0^{N_++N_-} Z
\end{equation}
where $\lambda_0$ is the bare fugacity which is related to the
particle chemical potential $\mu$ by $\lambda_0 = {\rm e}^{\mu}$.
We assume here that the chemical potential of plus and minus ions
is the same.
Using the definition of the exponential function,
${\rm e}^x= \sum_{N=0}^\infty x^N/N!$, the grand-canonical 
partition function can be written as
\begin{equation}
\label{part5}
Z_\lambda  = 
\int {\cal D} \rho_+  {\cal D} \rho_- {\cal D} \psi_+ {\cal D} \psi_-
\exp \left\{ -H[\rho_+, \rho_-,\psi_+, \psi_-]
\right\}
\end{equation}
with the field-theoretic action given by
\begin{eqnarray}
\label{ham1}
&& H [\rho_+, \rho_-,\psi_+, \psi_-]=  \nonumber \\ &&
\frac{\ell_B }{2} \int {\rm d} {\bf r}
 {\rm d} {\bf r}' [q\rho_+({\bf r})-q\rho_-({\bf r})+\sigma({\bf r}) ]
v({\bf r}-{\bf r}')  \nonumber \\ &&
[q\rho_+({\bf r}')-q\rho_-({\bf r}') + \sigma({\bf r}')] 
 \nonumber \\ && 
-\imath \int {\rm d} {\bf r} \left[ \psi_+({\bf r}) \rho_+ ({\bf r})
+ \psi_-({\bf r}) \rho_- ({\bf r}) \right]
 \nonumber  \\ &&
- \lambda 
\int {\rm d} {\bf r} \Omega({\bf r}) \left[
{\rm e}^{  - \imath \psi_+({\bf r}) }+  
{\rm e}^{- \imath \psi_-({\bf r}) } \right]
\end{eqnarray}
where we defined a rescaled fugacity as 
$\lambda = \lambda_0 {\rm e}^{q^2 \ell_B v(0)/2} /\lambda_t^3$.
As follows from the definition of the grand-canonical 
partition function, Eq.(\ref{part4}), the expectation
value of the total particle number is given by
\begin{eqnarray}
\label{norm1}
&& \langle N_+ + N_- \rangle = 
\lambda_0 \frac{\partial \ln Z_\lambda}{\partial \lambda_0}
= \lambda \frac{\partial \ln Z_\lambda}{\partial \lambda} =
\nonumber \\ &&
\lambda \int {\rm d} {\bf r} \Omega({\bf r}) \langle 
{\rm e}^{ - \imath \psi_+({\bf r})} +{\rm e}^{ - \imath \psi_-({\bf r}) } \rangle .
\end{eqnarray}
As can be be seen, the expectation value of the particle number,
$\langle N \rangle$,  is
independent of any multiplicative factor of the fugacity.
The standard way of saying this is that the chemical potential
is only defined up to a constant.

\subsection{Density description}

One type of approximations starts by eliminating the fluctuating
fields $\psi_+$ and $\psi_-$ from the problem. The simplest 
way of doing this is to replace the functional integral over these
fields by the value of the integrand at the extremum or optimal value.
The saddle point equations which determine the optimal values
of the fluctuating fields, $\psi_+^{SP}$ and  $\psi_-^{SP}$,
 are given by
\begin{equation} \label{SPeq}
\frac{\delta H [\rho_+, \rho_-,\psi_+, \psi_- ]}
{\delta \psi_+({\bf r})}=0 
\end{equation}
and with a similar equation for $\psi_-$.
Using $H$ as defined in Eq.(\ref{ham1}),
the solutions are
\begin{equation} \label{SPphi}
\psi_+^{SP}({\bf r}) = \imath \ln(\rho_+({\bf r}) / \lambda))
\end{equation}
(and a similar solution for $\psi_-$)
which, upon insertion into the Hamiltonian Eq.(\ref{ham1}), leads to 
the saddle-point Hamiltonian
\begin{eqnarray}
\label{ham2}
&& H_\lambda[\rho_+, \rho_-,\psi_+^{SP}, \psi_-^{SP}]=  \nonumber \\ &&
\frac{\ell_B }{2} \int {\rm d} {\bf r}
 {\rm d} {\bf r}' [q\rho_+({\bf r})-q\rho_-({\bf r})+\sigma({\bf r}) ]
v({\bf r}-{\bf r}')  \nonumber \\ &&
[q\rho_+({\bf r}')-q\rho_-({\bf r}') + \sigma({\bf r}')] 
 \nonumber \\ && 
+ \int {\rm d} {\bf r} \; \rho_+({\bf r}) \left(
\ln[ \rho_+({\bf r}) / \lambda] -1 \right).
\nonumber \\ &&
+ \int {\rm d} {\bf r} \; \rho_-({\bf r}) \left(
\ln[ \rho_-({\bf r}) / \lambda] -1 \right).
\end{eqnarray}
This expression is the starting point for local density functional theory\cite{Stevens}. 
The extremum of Hamiltonian Eq.(\ref{ham2}), determined by
the equation
\begin{equation} \label{SPeq2}
\frac{\delta H [\rho_+, \rho_-,\psi_+^{SP}, \psi_-^{SP} ]}
{\delta \rho_+({\bf r})}=0 
\end{equation}
(and with a similar equation for $\rho_-$),
 is the mean-field or Poisson-Boltzmann
density distribution. As is well-known, the mean-field equation 
(\ref{SPeq2}) can be solved in closed form only
for a few cases, and for more complicated situations one has to rely on additional
approximations. One popular way of doing this is to minimize the Hamiltonian
Eq.(\ref{ham2}) for a restricted functional form of the ion distribution, for example
a box distribution. In the appendix A1 we show how this can be done for a planar charged
wall (where comparison with the closed-form mean-field solution is possible). 
The draw back of box models is that the ion distribution is discontinuous and thus
rather artificial and for more complicated charge distributions the box geometry becomes
complicated. Another, more basic drawback of the density functional  Eq.(\ref{ham2}) is
that correlation effects are not included on the local level. Non-local
terms can in principle be taken into account by expanding around the 
saddle-point in Eq.(\ref{SPphi}), but the resulting functional is complicated.
In the next section we show that a variational approach is more useful when 
applied in the potential description.

\subsection{Potential description}

A more powerful variational formulation is possible by choosing 
a description in terms of the 
fluctuating potentials $\psi_+$ and $\psi_-$. 
The starting point is the observation that the integral over the density
fields $\rho_+$ and $\rho_-$ in Eq. (\ref{part5}) can be performed exactly. Introducing
the new field $\phi({\bf r}) = [\psi_+({\bf r}) - \psi_-({\bf r})]/2$, 
the partition function can be written exactly as
\begin{equation}
\label{part6}
Z_\lambda  = 
\int \frac{{\cal D} \phi}{Z_v}
\exp \left\{ -H[\phi]
\right\}
\end{equation}
with 
\begin{equation}
H[\phi] = \int {\rm d}{\bf r}\left\{ \frac{[\nabla \phi({\bf r})]^2}{8 \pi \ell_B q^2}
+ \frac{\imath \sigma({\bf r})  \phi({\bf r}) }{q}
-2 \lambda \Omega({\bf r}) \cos \phi({\bf r}) \right\}
\end{equation}
where we used the inverse of the Coulomb potential 
$v^{-1}({\bf r}) = - \nabla^2 \delta ({\bf r})/4 \pi$.
The symbol ${\cal Z}_v$ denotes the determinant of the Gaussian integral
\begin{equation} \label{partv}
Z_v = \int {\cal D} \phi 
\exp \left\{ -  \frac{1}{2q^2 \ell_B } \int {\rm d} {\bf r}
 {\rm d} {\bf r}' \phi({\bf r}) v^{-1}({\bf r}-{\bf r}')  \phi({\bf r}') 
\right\}
\end{equation}
and is a measure of the free energy of vacuum fluctuations. 

Next, we rescale all lengths
by the Gouy-Chapman length $\mu=1/(2 \pi q \ell_B \sigma_s)$ 
according to ${\bf r} = \mu \tilde{\bf r}$, where $\sigma_s$ 
denotes the surface charge density of the charged object under consideration.
We obtain the modified partition function 
\begin{equation}
\label{part7}
Z _\lambda  = 
\int  \frac{{\cal D} \phi}{Z_v}
\exp \left\{ -\tilde{H}[\phi]/\Xi
\right\}
\end{equation}
with the rescaled action
\begin{equation}
\label{ham3}
\tilde{H} [\phi]= \int \frac{ {\rm d} \tilde{\bf r}}{2\pi} 
\left[ \frac{1}{4} 
\left( \nabla \phi(\tilde{\bf r}) \right)^2
+ \imath \phi(\tilde{\bf r})
\tilde{\sigma} (\tilde{\bf r}) - \frac{ \Lambda \tilde\Omega( \tilde{\bf r})
\cos \phi(\tilde{\bf r})}{2}  \right]
\end{equation}
where we used the rescaled charge distribution
$\tilde{\sigma} ({\bf r}/\mu) = \mu \sigma({\bf r}) /\sigma_s$
and ion exclusion function $\tilde{\Omega} ({\bf r}/\mu) = \Omega({\bf r}) $.
The rescaled fugacity $\Lambda$ is defined by
\begin{equation} \label{Lambdef}
\Lambda = 8 \pi \lambda \mu^3 \Xi = \frac{2 \lambda}{\pi \sigma_s^2 \ell_B}
\end{equation}
and is determined by the normalization condition (in the case of fixed particle 
number)
\begin{eqnarray}
\label{norm2}
&& \langle N_+ + N_- \rangle 
= \Lambda \frac{\partial \ln Z_\Lambda}{\partial \Lambda} =
\nonumber \\ &&
\frac{\Lambda}{4 \pi \Xi} \int {\rm d} \tilde{\bf r} 
\tilde{\Omega}(\tilde{\bf r}) \langle 
\cos \phi (\tilde{\bf r}) \rangle .
\end{eqnarray}
As one can see, there are two parameters left in the
field-theoretic formulation, namely the 
 coupling parameter 
\begin{equation}
\Xi=2\pi \ell_B^2 \sigma_s q^3,
\end{equation}
which measures to which extent 
deviations from the saddle-point are important,
and the ion fugacity $\Lambda$.
Infinitely far from any charged object the potential
is constant and the concentration of minus and plus
ions is equal and given by
\begin{equation}
c_s = \frac{N_+}{V}=\frac{N_-}{V} = \frac{\Lambda}{8 \pi \Xi \mu^3}
\langle  \cos \phi \rangle
\end{equation}
which, using the definition of the screening length, $\kappa^{-1} =
(8\pi q^2 \ell_B c_s)^{-1/2}$ and the Gouy Chapman
length can be rewritten as 
\begin{equation} \label{Lambda}
\kappa^2 \mu^2 = \tilde{\kappa}^2 = \Lambda \langle  \cos \phi \rangle.
\end{equation}
The rescaled fugacity $\Lambda$ thus measures the ratio between the screening
length and the Gouy-Chapman length: For $\tilde{\kappa} \gg 1$ the screening
by salt is dominant, non-linear effects are unimportant and Debye-H\"uckel theory
is valid, for $\tilde{\kappa} \ll 1$ on the other hand non-linear effects are important
and Debye-H\"uckel theory fails. It is the latter case where our variational theory
becomes important.

\section{Variational Theory}

The action Eq.(\ref{ham3}) is too complicated to be solved
exactly. Previously, perturbative theories have been proposed
for the weak-coupling (small $\Xi$) regime in terms of a 
loop-expansion around the saddle-point solution (which corresponds
to the solution of the non-linear Poisson-Boltzmann equation)\cite{Netz1}
and for the strong-coupling (large $\Xi$) regime in terms
of a virial expansion\cite{Netz2,Moreira1,Moreira2}. 
Both expansions are of limited usefulness because of their bad 
convergence properties as one moves into the intermediate regime between the
two asymptotic limits\cite{Moreira3}. It is this point where variational
theories come in as a robust means to obtain reliable, useful results.

The standard Gibbs variational procedure consists in minimizing
the following variational free energy 
\begin{equation} \label{Gibbs}
{\cal F}_{Gibbs} = {\cal F}_0 -\langle \tilde{H}_0-\tilde{H} \rangle_0 /\Xi
\end{equation}
with respect to the variational parameters included in the variational 
Hamiltonian $\tilde{H}_0$. As discussed in detail in Appendix B the Gibbs
variational principle is equivalent to the first-order perturbational
variation on the observable which is conjugate to the variational parameter.
The perturbational variation scheme can be systematically improved
by going to higher order, the Gibbs variation cannot. Since we stay at the 
first-order level throughout this paper, we will use the Gibbs variational
notation, though one should keep in mind that this can be interpreted as
the first order result of a perturbational variational calculation.

\subsection{Most general Gaussian variation}

The most general, yet exactly tractable variational Hamiltonian, is a 
Gaussian  one
\begin{equation}
\tilde{H}_0[\phi] =
\frac{1}{2} \int_{\bf r,r'}
[ \phi({\bf r})- \imath \phi_0({\bf r})] v_0^{-1}({\bf r},{\bf r}') 
[ \phi({\bf r}') - \imath \phi_0({\bf r}')] 
\end{equation} 
where the Gaussian kernel $v_0^{-1}$ and the mean potential $\phi_0$ are
variational parameters. The variational free energy according to 
Eq. (\ref{Gibbs}) reads
\begin{eqnarray}
{\cal F}_{Gibbs} &=& -{1 \over 2} {\rm Tr} \log v_0
-\int_{\bf r}\frac{ (\nabla \phi_0)^2 }{8 \pi \Xi} \nonumber \\ &&
+  \int_{\bf r,  r'}\delta({\bf r} -{\bf r}') 
{\nabla_r \nabla_{r'} v_0 ({\bf r},{\bf r}') \over
{8 \pi \Xi}} \nonumber \\ &&
- \int_{\bf r} \frac{\tilde{\sigma}({\bf r}) \phi_0({\bf r})}
{2\pi \Xi} \nonumber \\ &&
-\frac{{\Lambda}}{4 \pi \Xi} \int_{\bf r} \tilde{\Omega} \ 
{\rm e}^{-\Xi v_0 ({\bf r}, {\bf r})/2} \cosh(\phi_0).
\end{eqnarray}
The variational equations are given by
$\delta {\cal F}_{Gibbs}  /\delta \phi_0({\bf r})=0$ and
$\delta {\cal F}_{Gibbs}  /\delta v_0^{-1}({\bf r},{\bf r}')=0$ and read

\begin{eqnarray}
&& {\nabla^2 \phi_0 \over 4 \pi}= {\tilde{\sigma} \over 2 \pi} + {\Lambda \over 4 \pi}
\Omega \ e^{-\Xi v_0 ({\bf r}, {\bf r}) /2} \sinh \phi_0 \\
&& -{\nabla ^2 v_0 ({\bf r}, {\bf r'}) \over 4 \pi \Xi} + 
{\Lambda \over 4 \pi }e^{-\Xi v_0 ({\bf r}, {\bf r}) /2} 
\cosh \phi_0 \ v_0({\bf r}, {\bf r'}) \nonumber \\
&&= \delta ({\bf r}- {\bf r'}) 
\end{eqnarray}

These self-consistent equations are quite complicated even for the simplest
charge distribution. In the following we will devise a simpler, one-parameter
variational scheme which
will lead -as a byproduct- to a very simple and
clear definition of the renormalized or effective charge of an object.

\subsection{Variational Charge Renormalization}

As we will show in the following, non-linear effects are captured in an essentially
exact way already by a much simpler variational form, which as the only variational
parameter contains a charge-renormalization factor. The variational Hamiltonian 
reads 
\begin{equation}
\label{ham4}
\tilde{H}_0 [\phi]= \int \frac{ {\rm d} \tilde{\bf r}}{2\pi} 
\left[ \frac{1}{4} 
\left( \nabla \phi(\tilde{\bf r}) \right)^2
+ \imath \eta(\tilde{\bf r}) \phi(\tilde{\bf r})
\tilde{\sigma} (\tilde{\bf r}) + \frac{ \tilde{\kappa}^2 
\phi^2(\tilde{\bf r})}{4}  \right]
\end{equation}
where the variational function $\eta({\bf r})$ multiplies the charge 
distribution and therefore corresponds to a spatially dependent 
charge renormalization factor. The Hamiltonian is quadratic in the 
fluctuating field and therefore all expectation values can be calculated
by using Green's function formalism. Also, the geometry-dependent factor
$\Omega({\bf r})$ is for the present formulation omitted and the salt
concentration
dependent term proportional to $\tilde{\kappa}^2$ has the same strength 
throughout the space. This is certainly a simplification, which however allows
to use a simple translationally invariant Green's function (future
elaborations should probe variational forms with a spatially
dependent mass term).
Using the Debye-H\"uckel potential
\begin{equation}
v_{\rm DH} ({\bf r}) = \frac{{\rm e}^{-\tilde{\kappa} r}}{r}
\end{equation}
and the potential distribution
\begin{equation}
\phi_0 ({\bf r})= \frac{1}{2\pi} \int_{\bf r'} 
\eta({\bf r}') \tilde{\sigma}({\bf r'}) 
v_{\rm DH}({\bf r} - {\bf r}')
\end{equation}
the variational Hamiltonian Eq.(\ref{ham4}) can be rewritten as
\begin{eqnarray} \label{ham5}
\tilde{H}_0[\phi] &=&
\frac{1}{2} \int_{\bf r,r'}
[ \phi({\bf r})+ \imath \phi_0({\bf r})] v_{\rm DH}^{-1}({\bf r},{\bf r}') 
[ \phi({\bf r}')+ \imath \phi_0({\bf r}')]  \nonumber \\ &&
+\frac{1}{4\pi} \int_{\bf r} \eta({\bf r}) \tilde{\sigma}({\bf r}) \phi_0({\bf r}).
\end{eqnarray} 
The variational free energy according to Eq.(\ref{Gibbs}) is easily calculated and reads
\begin{eqnarray} \label{Gibbs2}
{\cal F}_{Gibbs} &=&{\cal F}_{\rm DH}  + 
\frac{1}{4\pi \Xi} \int_{\bf r} (2-\eta({\bf r})) \tilde{\sigma}({\bf r}) \phi_0({\bf r})  \\
&& +\frac{\tilde{\kappa}^2}{8 \pi \Xi} \int_{\bf r} [ \phi_0^2 ({\bf r})-\Xi v_{\rm DH}(0)]
\nonumber \\ && -\frac{\tilde{\kappa}^2}{4 \pi \Xi}  \int_{\bf r} 
\tilde{\Omega}({\bf r})  \cosh[\phi_0({\bf r}) ] \nonumber
\end{eqnarray}
where we used the short-hand notation
\begin{eqnarray}
{\cal F}_{\rm DH} &=& 
-\ln \int {\cal D} \phi \ {\rm e}^{-\int_{{\bf r}, {\bf r}'} \phi({\bf r}) 
v_{\rm DH}^{-1} ({\bf r} - {\bf r}') \phi({\bf r}')/2 \Xi} \nonumber \\
&=& -{1 \over 2} \int {d^3 q \over (2 \pi) ^3} \log ({\bf q}^2 + \tilde{\kappa}^2)
\end{eqnarray}
and fixed the fugacity $\Lambda$ according to Eq.(\ref{Lambda}).
The variational equation follows by taking a functional derivative of the 
free energy with respect to the charge renormalization function,
$\delta {\cal F}_{Gibbs}  /\delta \eta({\bf r})=0$, and 
reads after a few algebraic manipulations
\begin{eqnarray} \label{Gibbs3} &&
\int_{\bf r} (1-\eta({\bf r})) \tilde{\sigma}({\bf r}) \phi_0({\bf r})=
\nonumber \\ &&
\frac{\tilde{\kappa}^2}{2}  \int_{\bf r}\phi_0({\bf r}) \left(  
\tilde{\Omega}({\bf r}) \sinh[\phi_0({\bf r}) ]- \phi_0 ({\bf r})\right).
\end{eqnarray}
This integral equation is to be solved by a suitably chosen function
$\eta({\bf r})$. Since in this paper we will be only concerned with
 situation where translational invariance holds, 
we will in the following section specialize the variational approach to the
homogeneous case.

\subsection{Homogeneous variational approach}

The homogeneous variational approach is obtained by choosing a charge renormalization which
is uniform in space, i.e. $\eta({\bf r}) = \eta_0$. We redefine the bare potential as
\begin{equation} \label{phi0}
\phi_0 ({\bf r})= \frac{1}{2\pi} \int_{\bf r'} 
 \tilde{\sigma}({\bf r'}) 
v_{\rm DH}({\bf r} - {\bf r}')
\end{equation}
and the Hamiltonian can be rewritten as
\begin{eqnarray}
\tilde{H}_0[\phi] &=&
\frac{1}{2} \int_{\bf r,r'}
[ \phi({\bf r})+ \imath \eta_0\phi_0({\bf r})] v_{\rm DH}^{-1}({\bf r},{\bf r}') 
[ \phi({\bf r}')+ \imath \eta_0 \phi_0({\bf r}')]  \nonumber \\ &&
+\frac{\eta_0^2}{4\pi} \int_{\bf r} \tilde{\sigma}({\bf r}) \phi_0({\bf r}).
\end{eqnarray} 
The variational free energy according to Eq.(\ref{Gibbs}) reads
\begin{eqnarray} \label{Gibbs4}
{\cal F}_{Gibbs} &=&{\cal F}_{\rm DH}  + 
\frac{1}{4\pi \Xi} \int_{\bf r} (2-\eta_0) \eta_0 \tilde{\sigma}({\bf r}) \phi_0({\bf r})  \\
&& +\frac{\tilde{\kappa}^2}{8 \pi \Xi} \int_{\bf r} 
[ \eta_0^2 \phi_0^2 ({\bf r})-\Xi v_{\rm DH}(0)]\nonumber \\ &&
-\frac{\tilde{\kappa}^2}{4 \pi \Xi}  \int_{\bf r} \tilde{\Omega}({\bf r})
\cosh[\eta_0 \phi_0({\bf r}) ] \nonumber
\end{eqnarray}
The variational equation follows by taking a derivative with respect to
the charge renormalization factor,
$\partial {\cal F}_{Gibbs}  /\partial \eta_0=0$, and is given by
\begin{eqnarray} \label{SCE}
&& 2 (1-\eta_0) \int_{\bf r} \tilde{\sigma}({\bf r}) \phi_0({\bf r})  \\
&& = \tilde{\kappa}^2 \int_{\bf r}  \phi_0({\bf r}) \left( \tilde{\Omega}({\bf r})
 \sinh[\eta_0 \phi_0({\bf r}) ] -\eta_0 \phi_0({\bf r}) \right) . \nonumber
\end{eqnarray}
In the following we will solve this self-consistent equation for the charge-renormalization 
factor $\eta_0$ in the planar, cylindrical, and spherical geometries.

\subsubsection{Planar case}
For the calculation of the uniformly charged planar wall we assume that salt is present
on both sides of the wall, the geometry factor 
therefore equals unity $\tilde{\Omega}({\bf r})=1$ (it is an easy exercise
to generalize the calculation to the case where salt is present only on one
side of the charged wall).
With the bare charge distribution $\sigma({\bf r} ) = \sigma_s \delta (z)$ the 
rescaled charge distribution becomes $\tilde{\sigma}(\tilde {\bf r}) = \delta (\tilde{z})$.
The bare potential is according to Eq.(\ref{phi0}) 
given by $\phi_0 (\tilde{z}) = {\rm e}^{-\tilde{\kappa} \tilde{z}}/
\tilde{\kappa}$. The integrals in Eq.(\ref{SCE}) can all be performed exactly, and the resulting 
equation becomes 
\begin{equation}
1-\frac{\eta_0}{2} -\frac{\tilde{\kappa}^2}{ \eta_0} 
\left( \cosh[\eta_0/ \tilde{\kappa}] -1 \right)=0.
\end{equation}
In the limit $\tilde{\kappa} \gg 1$ the solution is
\begin{equation}
\eta_0 \simeq 1-\frac{1}{24 \tilde{\kappa}^2}
\end{equation}
and therefore only weak charge-regulation takes place, as expected for a weakly
charged surface or high salt concentrations. In the opposite limit of a highly
charged surface or at low salt concentrations, for  $\tilde{\kappa} \ll 1$, the solution is
\begin{equation}
\eta_0 \simeq - \tilde{\kappa} \ln(\tilde{\kappa}) .
\end{equation}
These results can be directly compared with the Poisson-Boltzmann results derived explicitly
in Appendix A2. 
The quantity one wants to compare is the surface potential, since this is conjugate to
the surface charge density, i.e. the variational parameter, and one can reinterpret the 
Gibbs variational principle as a first-order perturbational variation on the 
expectation value of the surface potential (as is explained in more detail in Appendix B).
Within our scheme, the surface potential is defined by 
\[ \psi_0=\eta_0 \phi_0(\tilde{z}=0)\]
and is given in units of $k_BT$ and valency $q$.
For the planar case, we obtain 
$\psi_0= \eta_0/\tilde{\kappa}$, and therefore find for the surface potential
$\eta_0/\tilde{\kappa} = 1/\tilde{\kappa}$ at high salt ($\tilde{\kappa} \gg 1$) and
$\eta_0/\tilde{\kappa} = -\ln(\tilde{\kappa}) $ at low salt ($\tilde{\kappa} \ll 1$).
These are the surface potentials one obtains from the full non-linear
Poisson-Boltzmann theory (except a factor of two for the low-salt case). 
This shows that our variational procedure captures non-linear effects accurately
and justifies to consider more complicated geometries, where the Poisson-Boltzmann 
equation can only be solved numerically.

\subsubsection{Cylindrical case}

We assume a charged cylinder with radius $R$ and linear charge density $\tau$.
The surface charge density follows as $\sigma_s = \tau / (2 \pi R)$.
Defining the Gouy-Chapman length (like in the planar case) 
as $\mu=1/(2 \pi q \sigma_s)$, we obtain $\mu = R/(q \tau \ell_B)$. 
The rescaled cylinder radius,
$\tilde{R} = R/\mu= q \tau \ell_B$ is thus equal to the Manning parameter\cite{Manning}:
For $\tilde{R}>1$ one expects counterions to be bound to the cylinder even in the
limit of vanishing salt concentration, for $\tilde{R}< 1$ no counterions stay bound in 
that limit. In the following, we assume the cylinder to be penetrable for ions,
the geometry function is $\tilde{\Omega}(\tilde{r}) = 1$ for all values of $\tilde{r}$.
Using cylindrical coordinates, 
the potential $\phi_0$ as defined in Eq.(\ref{phi0}) satisfies the differential
equation
\[ -\frac{1}{\tilde{r}} \frac{\partial}{\partial \tilde{r}} \tilde{r} 
\frac{\partial}{\partial \tilde{r}} \phi_0(\tilde{r}) + 2  \delta(\tilde{r} - \tilde{R})
+\tilde{\kappa}^2 \phi_0(\tilde{r}) = 0 \]
with the boundary conditions $\phi_0(\tilde{R}-0)=\phi_0(\tilde{R}+0)$ and
$ \phi_0'(\tilde{R}+0) - \phi_0'(\tilde{R}-0) =2$. 
The complete solution is constructed from  the two solutions  of the homogeneous
differential equation,
$\phi_0(\tilde{r}) =K_0(\tilde{\kappa} \tilde{r})$ and 
$\phi_0(\tilde{r}) =I_0(\tilde{\kappa} \tilde{r})$,  and reads 
\begin{equation} \label{phicyl1}
\phi_0(\tilde{r}) = 2 \tilde{R } I_0(\tilde{\kappa} \tilde{R}) 
K_0(\tilde{\kappa} \tilde{r}) 
\end{equation}
for $\tilde{r} > \tilde{R}$ and
\begin{equation} \label{phicyl2}
\phi_0(\tilde{r}) = 2 \tilde{R } K_0(\tilde{\kappa} \tilde{R}) 
I_0(\tilde{\kappa} \tilde{r}) 
\end{equation}
for $\tilde{r} < \tilde{R}$.
The solution of the integral equation (\ref{SCE}) can be easily obtained
graphically 
by calculating $\eta_0$ as a function of $\tilde{\kappa} \tilde{R}$ 
and $\eta_0 / \tilde{\kappa}$ via numerical integration followed by
a parametric plot. In Fig. 1a we show the renormalization factor 
$\eta_0$ as a function of the Manning parameter $\tilde{R}$ for fixed values
of the rescaled inverse screening length $\kappa R =5, 1, 0.2, 0.01$ 
(thin solid lines from top to bottom). The thick solid line denotes the Manning limit
\begin{equation} \label{Manning}
\eta_0 = 1/\tilde{R} = 1/(\tau \ell_B q)
\end{equation}
which is very slowly approached as the salt concentration decreases\cite{Levin}.
Indeed, an asymptotic analysis of the self-consistent equation (\ref{SCE})
with the cylinder potential Eqs. (\ref{phicyl1},\ref{phicyl2}) 
gives for $\kappa R \ll 1$
\begin{equation} \label{asym1}
\eta_0 \simeq \frac{1}{\tilde{R}} \left( 1+ \frac{ \ln \tilde{R}}{2 \ln (1/\kappa R)}
\right)
\end{equation}
which shows that the approach to the Manning limit is logarithmically slow. 
In the unrescaled units, the effective cylinder charge density is
\begin{equation}
\tau_{\rm eff} = \tau \eta_0 \simeq \frac{1}{\ell_B q} 
 \left( 1+ \frac{ \ln \tau \ell_B q}{2 \ln (1/\kappa R)}
\right)
\end{equation}
In Fig.1b we show 
the data for $\eta_0$ for fixed Manning parameter $\tilde{R} = 3 $
(which roughly corresponds to a fully charged vinyl-based polyelectrolyte
with a distance of $0.254$ $nm$ between charges along the backbone) as a function
of $\kappa R$ together with the asymptotic prediction Eq.(\ref{asym1}) as a
broken line in a semi-logarithmic plot and in Fig.1c as a function of 
$-1/\ln(\kappa R)$. 
It is seen that the asymptotic law is valid for low salt concentrations
satisfying $\kappa R < 0.1$. For large salt concentrations, asymptotic analysis
predicts that the charge-renormalization vanishes according to 
\begin{equation} \label{asym2}
\eta_0 \simeq 1-\frac{1}{24\tilde{\kappa}^2} 
\end{equation}
(identical to the result for the planar case)
which is shown in Fig.1b as a dotted line.

\begin{figure*}
  \begin{center}
    \epsfig{file=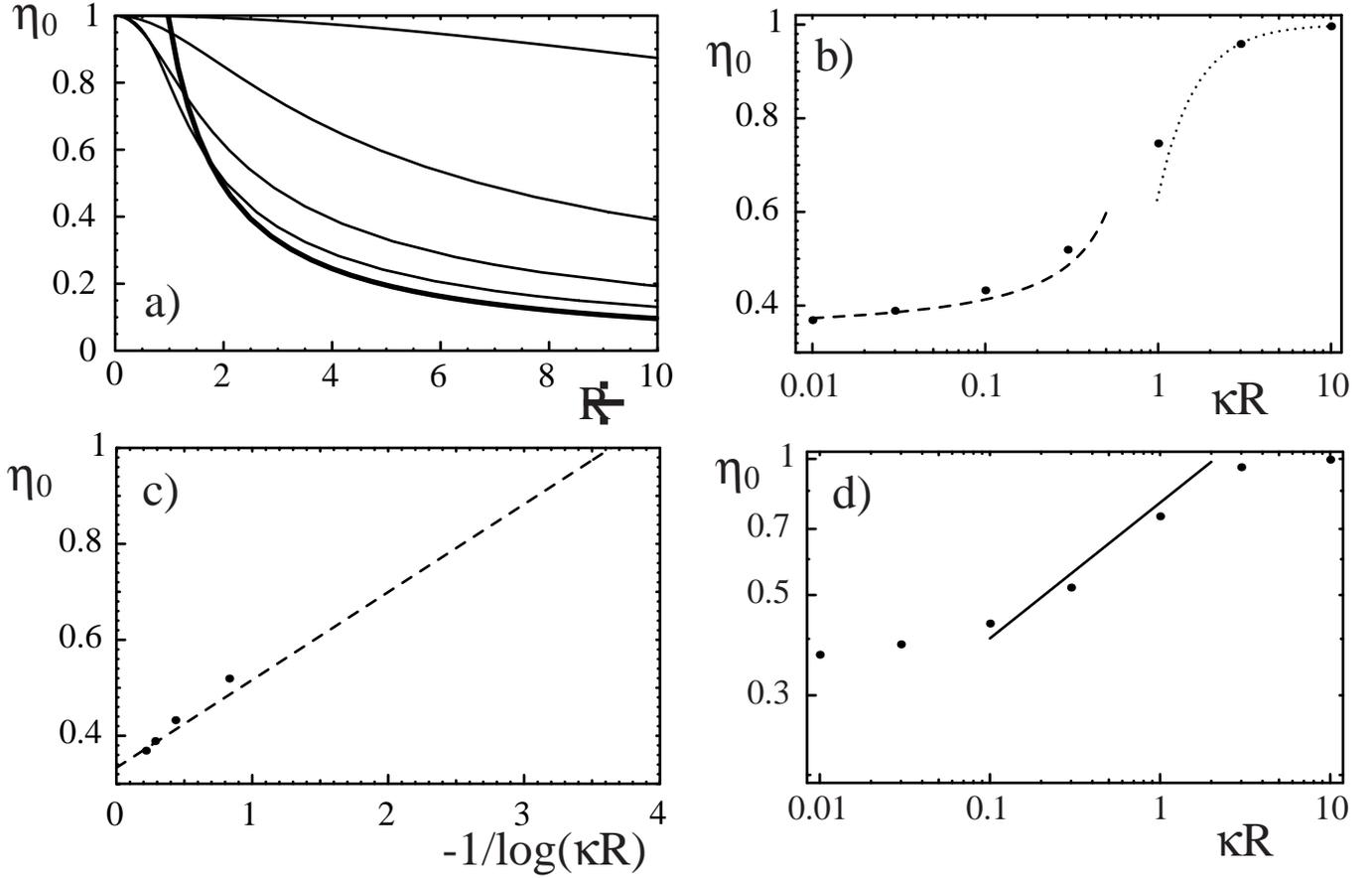, width=\textwidth}
  \end{center}
  \caption{a) Charge renormalization factor $\eta_0$ for a charged cylinder
  of radius $R$ 
 as a function of the Manning parameter $\tilde{R}=q \ell_B \tau$ for fixed values
of the rescaled inverse screening length $\kappa R =5, 1, 0.2, 0.01$ 
(thin solid lines from top to bottom). The thick solid line denotes the Manning
limit Eq.(\ref{Manning}). 
b-d) Results for fixed Manning parameter $\tilde{R}=3$ as a function of the rescaled radius
compared with the asymptotic low-salt prediction Eq.(\ref{asym1}) (broken line) and
the asymptotic high-salt prediction Eq.(\ref{asym2}) (dotted line). 
The solid line in d) corresponds to the power-law $\eta_0 \sim \kappa^{0.3}$. }
  \label{fig1}
\end{figure*}

\subsubsection{Nonlinear electrostatic persistence length}

The electrostatic contribution to the bending rigidity has been calculated
on the linear level\cite{Odijk,Skolnick} but also on the non-linear level using the 
full Poisson-Boltzmann equation for a bent cylinder\cite{LeBret,Fixman}.
Here we re-address the relevance of  non-linear effects for the persistence
length of highly charged polyelectrolytes.
In Fig.1d we plot the charge renormalization factor on double logarithmic scales.
Over roughly a decade in $\kappa R$, which corresponds to two decades in the salt
concentration, the charge renormalization factor can be approximately described
by a power law $\eta_0 \sim \kappa^{\alpha}$ with $\alpha \approx 0.3 $ (shown 
as a solid line). This range of salt concentrations is quite relevant 
for experiments\cite{Forster}, since it has been reported that the experimentally
measured salt-concentration dependence of the persistence length 
deviates from the Odijk-Skolnick-Fixman prediction, according to which 
\begin{equation} \label{OSF}
\ell_{\rm OSF} = \frac{\ell_B \tau^2}{4 \kappa^\beta}
\end{equation}
with $\beta =2$\cite{Odijk,Skolnick}. 
Replacing the bare line charge density $\tau$ in Eq.(\ref{OSF})
by the effective charge density 
\begin{equation}
\tau_{\rm eff} = \eta_0 \tau \sim \kappa^\alpha \tau
\end{equation}
(which is permissible since to leading order in an expansion for large bending radii
modifications of the charge renormalization due to polymer bending are irrelevant)
one obtains a modified exponent $\beta = 2-2 \alpha \approx 1.4$ which is closer
to the experimentally measured exponents\cite{Forster}. Other effects that have been
shown to reduce the exponent $\beta$ are relevant for weakly charge chains which 
are crumpled at small length scales\cite{Joanny,Sim2,Micka,Thirumalai} although recent
simulations using linear Debye-H\"uckel potentials\cite{Ralf,Boris}
as well as general Gaussian variational calculations\cite{Netz4}
 tend to confirm the original
 Odijk-Skolnick-Fixman prediction with $\beta=2$.
It might therefore turn out that the true explanation for the discrepancy between
experimental and theoretical predictions for the electrostatic persistence length
is due to non-linear effects which had been neglected in most theoretical treatments
but are of course omnipresent in experiments. Needless to say, we expect modification
of a whole number of other scaling relations for polyelectrolytes
which depend on the line charge density $\tau$.

\subsubsection{Spherical Case}

We now treat a sphere of charge $Z$ and radius $R$, leading to a surface charge
density $\sigma_s = Z/(4 \pi R^2)$ and Gouy-Chapman length $\mu = 1/(2 \pi q \ell_B \sigma_s)$.
The rescaled radius is given by $\tilde{R} = R/\mu = q \ell_B Z/(2 R)$. 

Using spherical coordinates, 
the potential $\phi_0$ as defined in Eq.(\ref{phi0}) satisfies the differential
equation
\[ -\frac{1}{\tilde{r}^2} \frac{\partial}{\partial \tilde{r}} \tilde{r}^2 
\frac{\partial}{\partial \tilde{r}} \phi_0(\tilde{r}) + 2  \delta(\tilde{r} - \tilde{R})
+\tilde{\kappa}^2 \phi_0(\tilde{r}) = 0 \]
with the boundary conditions $\phi_0(\tilde{R}-0)=\phi_0(\tilde{R}+0)$ and
$ \phi_0'(\tilde{R}+0) - \phi_0'(\tilde{R}-0) =2$. 
The two solutions  of the homogeneous
differential equation are
$\phi_0(\tilde{r}) ={\rm e}^{\tilde{\kappa} \tilde{r}}/\tilde{r}$ and 
$\phi_0(\tilde{r}) ={\rm e}^{-\tilde{\kappa} \tilde{r}}/\tilde{r}$,
from which we obtain the solutions
\begin{equation} \label{phisph1}
\phi_0(\tilde{r}) = 2 \tilde{R } \sinh (\tilde{\kappa} \tilde{R}) 
{\rm e}^{-\tilde{\kappa} \tilde{r}}/(\tilde{\kappa} \tilde{r}) 
\end{equation}
for $\tilde{r} > \tilde{R}$ and
\begin{equation} \label{phisph2}
\phi_0(\tilde{r}) = 2 \tilde{R } {\rm e}^{-\tilde{\kappa} \tilde{R}}
\sinh (\tilde{\kappa} \tilde{r}) /(\tilde{\kappa} \tilde{r}) 
\end{equation}
for $\tilde{r} < \tilde{R}$ where we have again assumed that salt ions are present
everywhere, i.e.  $\tilde{\Omega}(\tilde{r}) = 1$ throughout space.

\begin{figure}
  \begin{center}
    \epsfig{file=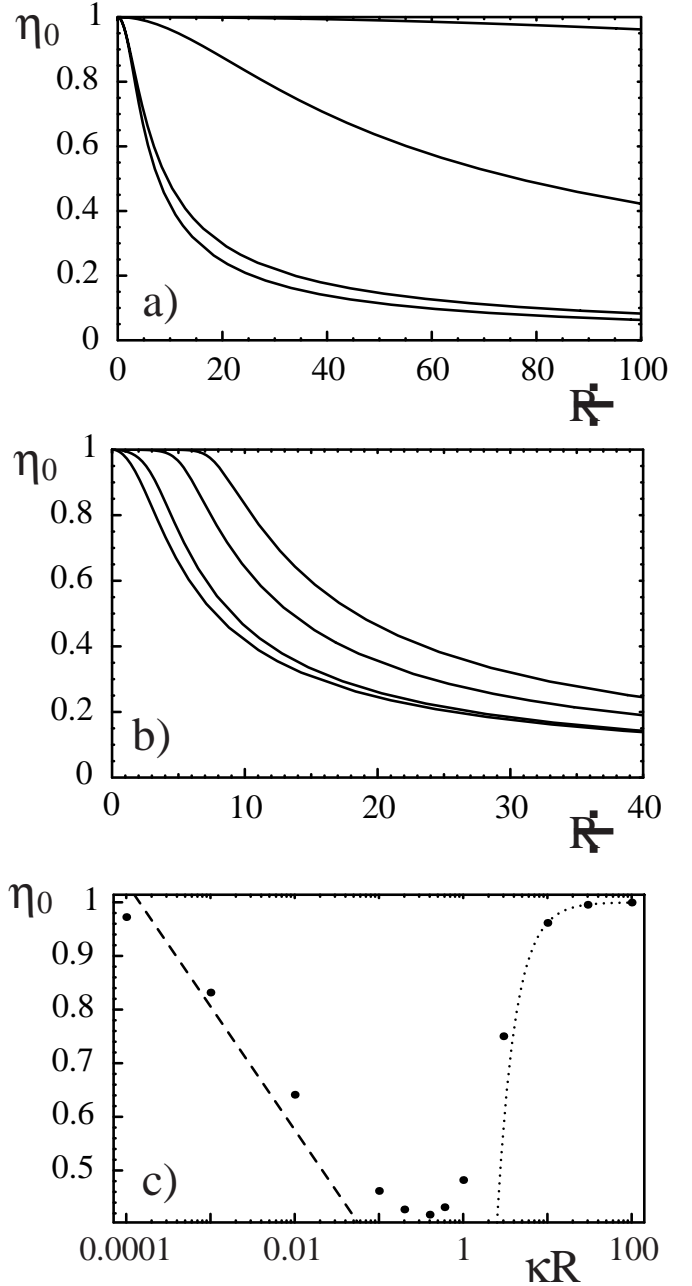, width=88mm}
  \end{center}
  \caption{Results for the charge renormalization factor $\eta_0$ for a sphere
  for rescaled inverse screening lengths a)  
$\kappa R = 100, 10, 1, 0.4 $ (from top to bottom) and b) 
for $\kappa R = 0.4,, 0.1, 0.01, 0.001$ (from bottom to top).
c) Charge renormalization factor for fixed $\tilde{R}=R/\mu = 10$ 
together with the asymptotic prediction Eq.(\ref{asym4}) (broken line)
 and the prediction Eq.(\ref{asym3})  (dotted line).
}
  \label{fig2}
\end{figure}

In Fig.2a we show parametric plots of the charge renormalization factor for 
$\kappa R = 100, 10, 1, 0.4 $ (from top to bottom) and in Fig.2b 
for $\kappa R = 0.4,, 0.1, 0.01, 0.001$ (from bottom to top). In contrast to the 
cylindrical case, charge renormalization is most pronounced at intermediate salt concentrations
corresponding to a screening length which roughly equals the sphere radius\cite{Belloni}. 
For large salt concentrations, asymptotic analysis
predicts that the charge renormalization vanishes according to 
\begin{equation} \label{asym3}
\eta_0 \simeq 1-\frac{1}{24\tilde{\kappa}^2} 
\end{equation}
in agreement with the planar and cylindrical geometries.
For low salt concentrations,
 an asymptotic analysis of the self-consistent equation (\ref{SCE})
with the spherical potential Eqs. (\ref{phisph1},\ref{phisph2}) 
gives for $\kappa R \ll 1$
\begin{equation} \label{asym4}
\eta_0 \simeq \frac{1}{\tilde{2R}}  \ln \left( \frac{1}{\tilde{R} \tilde{\kappa}^2 }\right).
\end{equation}
In unrescaled units this leads to an effective sphere charge
\begin{equation}
Z_{\rm eff} = \eta_0 Z \simeq \frac{R}{q \ell_B} \ln
 \left( \frac{q \ell_B Z}{R^3 \kappa^2} \right)
 \end{equation}
 which only depends very weakly on the bare sphere charge (in agreement
 with previous qualitative arguments\cite{Pincus}). 
 In Fig.2c we compare data for a sphere with rescaled radius $\tilde{R} = R/\mu=10$ 
 with the asymptotic low-salt prediction Eq.(\ref{asym4}) (broken line)
 and the high-salt prediction Eq.(\ref{asym3})  (dotted line). Note that the asymptotic
 low-salt prediction is only valid in a finite salt-concentration limit: for 
 very low salt concentration, the counterions are entropically driven away from the
 sphere and the bare charge is recovered. This might be detectable in 
 light-scattering experiments on de-salted charged-colloid suspensions.

\section{Conclusions}

We have introduced the framework for a general variational technique
which allows to capture non-linear effects of counterion distributions 
around highly charged objects in an essentially exact way. 
The scheme builds on the field-theoretic formulation of fluctuating
counterions and uses the effective surface charge as a variational 
parameter. Since we use a Gaussian variational Hamiltonian, the resulting
interactions between charged particles are of the Debye-H\"uckel type,
but with the bare charges replaced by effective ones. This scheme
is thus very close to the treatment by Kjellander et al.\cite{Kjellander}, 
where a similar
procedure was used within an integral-equation scheme.
We show how in principle to systematically improve the leading term results
by using perturbational variational methods.
We tested our approach for the planar geometry, where comparison with
the analytical solution of the non-linear Poisson-Boltzmann equation 
is possible, and then presented a detailed analysis of the cylindrical 
and spherical case. In the future, we will apply these methods
to the interaction between charged bodies (including the experimentally
relevant case of a charged cylinder adsorbing on a charge plane\cite{fleck}),
soft (fluctuating) charged polymers and membranes. 
We also plan to investigate 
higher-order perturbation variational terms which will allow to include 
effects at high coupling parameter $\Xi$ where deviations from the 
mean-field like behavior are important.


\appendix

\section{Appendix: Charged planar wall}

In this section we will treat the single charged wall immersed in a salt solution using 
a box model and using the Poisson Boltzmann equation.

\subsection{Box model}
The mean-field Hamiltonian in terms of the ionic densities, Eq.(\ref{ham2}),
 can be used as a starting
point for a simplified variational treatment using a box model, where the true 
ionic density distribution is replaced by a step profile.
The fugacity can be fixed by the ionic density in the bulk where we assume the 
electrostatic interactions to vanish. Minimization Eq.(\ref{ham2}) 
with respect to the density of plus ions yields
\begin{equation}
\frac{\delta H}{\delta \rho_+} = \ln(\rho_+/\lambda) = 0
\end{equation}
and thus $\rho_+ = \lambda$. We conclude that the fugacity is given by the
ionic density in the bulk reservoir, i.e. $\lambda = \rho_+^b$. 

As a box model we assume a negatively charged planar wall of surface charge density
$\sigma_s$ which is neutralized by a uniformly charged box of width $d$. 
We assume that the wall charge is neutralized by an excess of positive ions, 
i.e. the density of negative ions stays constant. The electric field distribution
thus is
\begin{equation}
\frac{eE}{k_BT} = 4 \pi \ell_B \sigma_s \frac{d-z}{d}
\end{equation}
and the electrostatic energy (per unit area and per $k_BT$) follows as
\begin{eqnarray}
U &=& \frac{\varepsilon}{2k_BT} \int {\rm d}z E^2 =
\frac{1}{8 \pi \ell_B}   \int {\rm d}z \left( \frac{e E}{k_BT} \right)^2
\nonumber \\ && = 2 \pi \ell_B \sigma_s^2 d /3.
\end{eqnarray}
The excess entropy of the layer of condensed counterions is 
\begin{equation}
-S= d\left\{ \left( \rho_+^b + \frac{\sigma}{qd} \right) \left( \log \left[
1 + \frac{\sigma_s}{qd \rho_+^b} \right]-1\right) +\rho_+^b \right\}.
\end{equation}
Minimization of the free energy $F = U-S$ leads for high salt concentrations
(for $\kappa \mu =\tilde{\kappa} \gg 1$) 
to the optimal layer thickness
\begin{equation}
d^2 \simeq \frac{3}{4 \pi \ell_B q^2 \rho_+^b} = 6 \kappa^{-2}
\end{equation}
which is proportional to the screening length $\kappa^{-1}$
(as expected)
and for low salt concentrations (for $\kappa \mu \ll 1$) to
\begin{equation}
d \simeq \frac{3}{2 \pi \ell_B q \sigma_s} = 3 \mu
\end{equation}
and thus to a layer thickness which differs from the Gouy Chapman 
length $\mu$ only by a numerical factor. 
The surface free energies are given by
\begin{equation}
F \simeq \frac{2 \sqrt{6} \sigma_s}{3 q \kappa \mu } 
\end{equation}
for high salt and 
\begin{equation}
F \simeq \frac{2  \sigma_s}{ q} \log \left( \frac{1}{\kappa \mu} \right). 
\end{equation}
for low salt.

\subsection{Poisson-Boltzmann equation}

The Poisson-Boltzmann equation follows from minimizing the Hamiltonian 
Eq.(\ref{ham3}) with respect to the potential distribution.
Defining the real electrostatic potential $\psi = \imath \phi$
the saddle point equation can be written as
\begin{equation}
\psi''(z) + 2 \delta(z) - \tilde{\kappa}^2 \sinh \psi =0.
\end{equation}
The solution is 
\begin{equation}
\psi(z) = 2 \log\left( \frac{1-\gamma {\rm e}^{-\kappa z}}
{1 + \gamma {\rm e}^{-\kappa z}} \right) 
\end{equation}
where the constant $\gamma$ is determined by the boundary conditions.
In the case when the charged wall is impenetrable to ions, the 
boundary condition is $\psi'(0) = -2$, which leads to the equation
$\gamma^2 -1 = 2 \gamma \tilde{\kappa}$. For low salt concentrations the
constant follows as $\gamma \simeq -1 + \kappa$ and the 
surface potential is given by
\begin{equation}
\psi_0 \simeq 2 \ln(1/\tilde{\kappa}).
\end{equation}
The surface free energy in this limit is $F = \psi_0 \sigma_s /(2 q) = 
(\sigma_s/q) \ln(1/\tilde{\kappa})$ and thus differs from the box-model calculation
only by a prefactor.
For high salt concentrations the constant is $\gamma = -1/(2 \tilde{\kappa})$
and the surface potential is 
\begin{equation}
\psi_0 \simeq 2 /\kappa.
\end{equation}
The surface free energy is $F = \sigma_s/(\tilde{\kappa} q)$
and is also reproduced by the box-model calculation.
In the case when the surface is penetrable to ions, as is used
in our variational calculations, the surface potential in the low salt
limit is not modified to leading order,
\begin{equation}
\psi_0 \simeq 2 \ln(1/\tilde{\kappa}),
\end{equation}
while in the high salt limit one obtains
\begin{equation}
\psi_0 \simeq 1 /\kappa.
\end{equation}

\section{Appendix: Connection between the Gibbs Variational Principle 
and Perturbative Variation}

Let us start by considering the free energy 
\begin{equation}
{\cal F} = -\ln \int {\cal D} \phi {\rm e}^{- {\cal H}}
\end{equation}
where ${\cal H}$ depends on the fluctuating field $\phi$.
We introduce a variational Hamiltonian ${\cal H}_0({\bf x})$ 
which depends on the fluctuating field $\phi$ and on
the vector ${\bf x}$, the components of which
may act as variational parameters and/or as generating fields
to calculate observables. Adding and subtracting the 
variational Hamiltonian to the action, the free energy can be written as
\begin{equation}
{\cal F} = -\ln \int {\cal D} \phi {\rm e}^{- {\cal H}_0({\bf x})}-
\ln \left[ \frac{ \int {\cal D} \phi {\rm e}^{- {\cal H}_0({\bf x})+ \zeta Y}}
{\int {\cal D} \phi {\rm e}^{- {\cal H}_0({\bf x})}} \right] 
\end{equation}
where we defined 
\begin{equation}
Y = {\cal H}_0({\bf x})-{\cal H}
\end{equation}
and introduced a dummy parameter $\zeta$ which is used to count
orders in a perturbation series (and will eventually be set to unity).
The free energy can now be expanded in cumulants as
\begin{equation}
{\cal F}_n = {\cal F}_0({\bf x}) -\sum_{m=1}^n \frac{\zeta^m}{m!} \left\langle
Y^m \right\rangle_0^C
\end{equation}
where ${\cal F}_0({\bf x}) $ denotes the free energy of the 
variational Hamiltonian
\begin{equation}
{\cal F}_0({\bf x}) = -\ln \int {\cal D} \phi {\rm e}^{- {\cal H}_0({\bf x})}
\end{equation}
and $\left\langle \ldots \right\rangle_0^C$ denotes the cumulant average
with respect to the variational Hamiltonian. Clearly, the exact free energy is approached
in the infinite limit of the series, i.e. ${\cal F} = {\cal F}_{n \rightarrow \infty}$.
The standard perturbation technique is often slowly converging for problems of interest. 
Variational techniques typically perform much better. The standard Gibbs-variational 
technique exploits the fact that the first order perturbation result
${\cal F}_1$ is always larger than the true free energy, ${\cal F}_1 \geq {\cal F}$.
The approximate free energy is therefore given by minimizing ${\cal F}_1$ with respect
to the set of variational parameters,
\begin{equation} \label{Gibbsfree}
{\cal F}_{\rm Gibbs} = \min_{\bf x} {\cal F}_1({\bf x}).
\end{equation}
However, it is not easy to systematically improve this result (since the property
of the perturbative free energy to be  bound by the true free energy
is lost at higher order), other than by choosing 
a more refined variational Hamiltonian, which is often not possible. The perturbational
variation, which has been introduced by Edwards for the calculation
of the scaling behavior of self-avoiding polymers\cite{Edwards},
 consists in requiring the perturbative free energy of any order to be equal 
to the zeroth order free energy,
\begin{equation} \label{pertfree}
{\cal F}_0 ({\bf x})= {\cal F}_n({\bf x})
\end{equation}
which is equivalent to the condition
\begin{equation} \label{pert1}
\sum_{m=1}^n \frac{\zeta^m}{m!} \left\langle
Y^m \right\rangle_0^C =0.
\end{equation}
This equation has to be solved by fixing the set of variational parameters ${\bf x}$.
To do this, it is convenient to expand the parameters in a 
power series ${\bf x} = \sum_{i=1}^{n} \zeta^{i-1} {\bf  x}^{(i)}$ and to solve 
Eq.(\ref{pert1}) term by term. (If the variational parameter is not expanded in a series
but Eq.(\ref{pert1}) is solved directly, one recovers the naive perturbation result. 
The hope is therefore that the expansion of the variational parameters improves the 
convergence behavior of the series for the free energy.) One notes that the prediction for the 
free energy from the Gibbs variational approach, Eq.(\ref{Gibbsfree}), and the
perturbational variation, Eq.(\ref{pertfree}), are not equal, not even at the leading
order. It is a priori not clear which approximation closer to the exact free energy.

Often one is not interested in the free energy itself but in a certain observable.
We define an arbitrary observable $X_j$ as being  conjugate  to the 
field  $x_j$ according to
\begin{equation}
X_j = \frac{ \partial {\cal H}_0({\bf x}) }{\partial x_j}.
\end{equation}
The first few terms of the expansion of the expectation value of 
$X_j $ in powers of $\zeta$ are
\begin{eqnarray} \label{series}
&& \langle X_j \rangle = \langle X_j \rangle_0 + \zeta
\left[  \langle X_j Y \rangle_0 -\langle X_j \rangle_0 \langle Y \rangle_0 \right]
 \\ 
&+&\frac{\zeta^2}{2} \left[ \langle X_j Y^2 \rangle_0 
-2 \langle X_j Y \rangle_0 \langle Y \rangle_0
-\langle X_j  \rangle_0 \langle Y^2 \rangle_0
+2\langle X_j  \rangle_0 \langle  Y \rangle_0^2 \right].\nonumber
\end{eqnarray}
As before, 
the perturbative variation consists of demanding that the corrections to the
leading contribution of the expectation value vanish, i.e.
\begin{equation} \label{condition}
\langle X_j \rangle = \langle X_j \rangle _0 ,
\end{equation}
which can be solved by adjusting one or more of
the variational parameters ${\bf x}$.

A quite instructive connection exists between the Gibbs variational method
and the perturbative variation: In the case when the Gibbs variational parameter
is the field $x_j$ conjugate to the observable $X_j$, the Gibbs variational
equation and the first order perturbative variation Eq.(\ref{condition})
are identical. This means that the higher order perturbative variational method
on the expectation value can be viewed as a generalization of the Gibbs variational
technique.
One might hope to approximate the true free energy better,
though the true free energy ceases to be a strict lower bound for the perturbational
free energy expression.

Some comments are in order:

i) In principle, the calculated observable in equation (\ref{series}) 
does not need to be conjugate to the variational parameter. In any case,
it is clear that in principle the observable can be calculated to any
wanted precision by taking enough terms in the perturbation-variational
treatment.

ii) Several expectation values can be calculated, for which an equal number of 
variational parameters are needed to solve the uniform expansion condition.

v) It is not clear that equation (\ref{condition}) always has a solution.
In that case one could try to find the best estimate to the solution.

\bibliographystyle{}

\begin{thebibliography}{10}

\bibitem{Holm}
{\em Electrostatic Effects in Soft Matter and Biophysics},
C. Holm, P. Kekicheff, und R. Podgornik (Hrgb.), 
Kluwer Academic Publishers, Dordrecht (2001).

\bibitem{gouy}
G. Gouy, J. de Phys. {\bf IX},  457  (1910);
D.~L. Chapman, Phil. Mag. {\bf 25},  475  (1913).
For cylindrical geometry a solution in terms of an
infinite series has been obtained, C.A. Tracy and H. Widom,
Physica A {\bf 244}, 402 (1997).

\bibitem{Manning2}
G.S. Manning, Ber. Bunsenges. Phys. Chem. {\bf 100}, 909 (1996);
G.S. Manning and J. Ray, J. Biomolecular Structure and Dynamics 
{\bf 16}, 461 (1998).

\bibitem{Pincus}
S. Alexander, P.M. Chaikin, P. Grant, G.J. Morales, P. Pincus, D. Hone,
J. Chem. Phys. {\bf 80}, 5776 (1984).

\bibitem{Manning}
G.S. Manning,  J. Chem. Phys. {\bf 51},  924, 934 (1969).

\bibitem{Belloni}
L. Belloni, Coll. Surf. A {\bf 140}, 227 (1998).

\bibitem{Levin}
P.S. Kuhn, Y. Levin, and M.C. Barbosa, Macromolecules {\bf 31}, 8347 (1998).

\bibitem{Deserno}
M. Deserno, C. Helm, and S. May, Macromolecules {\bf 33}, 199 (2000).

\bibitem{Odijk}
T. Odijk, { J. Polym. Sci., Polym. Phys.. Ed.} {\bf 15}, 477 (1977);
{ Polymer} {\bf 19} 989 (1978).

\bibitem{Skolnick}
J. Skolnick and M. Fixman, Macromolecules {\bf 10} 944 (1977).

\bibitem{Joanny}
J.-L. Barrat,  J.-F. Joanny, { Europhys. Lett.} {\bf 24} 333 (1993).

\bibitem{Forster}
S. F\"orster and M. Schmidt,  Advances in Polymer Science, Vol. 120
(Springer Verlag Berlin Heidelberg 1995).

\bibitem{Stevens}
M.J. Stevens and M.O. Robbins, Europhys. Lett. {\bf 12}, 81 (1990).

\bibitem{Netz1}
R.R. Netz and H. Orland, Eur. Phys. J. E {\bf 1}, 203 (2000).

\bibitem{Netz2}
R.R. Netz,  Eur. Phys. J. E {\bf 5}, 557 (2001).

\bibitem{Moreira1}
A.G. Moreira and R.R. Netz, Europhys. Lett. {\bf 52}, 705 (2000).

\bibitem{Moreira2}
A.G. Moreira and R.R. Netz, Phys. Rev. Lett.  {\bf 87}, 078301 (2001).

\bibitem{Moreira3}
A.G. Moreira and R.R. Netz,  Eur. Phys. J. E {\bf 8}, 33 (2002).

\bibitem{LeBret}
M. Le Bret, J. Chem. Phys. {\bf 76}, 6243 (1982).

\bibitem{Fixman}
M. Fixman, J. Chem. Phys. {\bf 76}, 6346 (1982).

\bibitem{Sim2}
C. Seidel, H. Schlacken, and I. M\"uller,
Macromol. Theory Simul. {\bf 3}, 333 (1994).

\bibitem{Micka}
U. Micka and K. Kremer, Phys. Rev. E {\bf 54}, 2653 (1996).

\bibitem{Thirumalai}
B.-Y. Ha and D. Thirumalai, J. Chem. Phys. {\bf 110}, 7533 (1999).

\bibitem{Ralf}
R. Everaers, A. Milchev, and V. Yamakov, Eur. Phys. J. E {\bf 8}, 3 (2002).

\bibitem{Boris}
T.T. Nguyen and B.I. Shklovskii, Phys. Rev. E {\bf 66}, 021801 (2002).

\bibitem{Netz4}
R. R. Netz, H. Orland, { Eur. Phys. J. B} {\bf 8} 81 (1999).

\bibitem{Kjellander}
R. Kjellander, J. Chem. Phys. {\bf 99}, 10392 (1995).

\bibitem{fleck}
C. Fleck, R.R. Netz, H.H. von Gr\"unberg, 
Biophys. J. {\bf 82}, 76 (2002).

\bibitem{Edwards}
 {\em The Theory of Polymer Dynamics}, M. Doi and S.F. Edwards
(Oxford University Press, Oxford, 1986).

\end{thebibliography}

\end{document}